\documentclass[12pt]{article}

\usepackage{siunitx}
\usepackage{pdflscape}
\usepackage{changepage}
\usepackage{tabularx}
\usepackage{setspace}
\usepackage{substr}\usepackage{lipsum}\usepackage{placeins}
\usepackage{subfigure}
\usepackage{afterpage}
\newcommand{\m}{\texttt{Contact-mmc}}

\usepackage{pifont}
\newcommand{\xmark}{\ding{55}}%

\usepackage{caption}
\usepackage{longtable}
\usepackage{tikz}
\usepackage{tikz-3dplot}
\usepackage{pdfpages}
\usetikzlibrary{arrows}
\usetikzlibrary{calc}
\usepackage{mathrsfs}
\usetikzlibrary{calc,trees,positioning,arrows,chains,shapes.geometric,%
    decorations.pathreplacing,decorations.pathmorphing,shapes,%
    matrix,shapes.symbols}

\tikzset{
>=stealth',
  punktchain/.style={
    rectangle,
    rounded corners,
    draw=black, very thick,
    text width=10em,
    minimum height=3em,
    text centered,
    on chain},
  line/.style={draw, thick, <-},
  element/.style={
    tape,
    top color=white,
    bottom color=blue!50!black!60!,
    minimum width=8em,
    draw=blue!40!black!90, very thick,
    text width=10em,
    minimum height=3.5em,
    text centered,
    on chain},
  every join/.style={->, thick,shorten >=1pt},
  decoration={brace},
  tuborg/.style={decorate},
  tubnode/.style={midway, right=2pt},
}
\makeatletter
\pgfdeclareshape{datastore}{
  \inheritsavedanchors[from=rectangle]
  \inheritanchorborder[from=rectangle]
  \inheritanchor[from=rectangle]{center}
  \inheritanchor[from=rectangle]{base}
  \inheritanchor[from=rectangle]{north}
  \inheritanchor[from=rectangle]{north east}
  \inheritanchor[from=rectangle]{east}
  \inheritanchor[from=rectangle]{south east}
  \inheritanchor[from=rectangle]{south}
  \inheritanchor[from=rectangle]{south west}
  \inheritanchor[from=rectangle]{west}
  \inheritanchor[from=rectangle]{north west}
  \backgroundpath{
    \southwest \pgf@xa=\pgf@x \pgf@ya=\pgf@y
    \northeast \pgf@xb=\pgf@x \pgf@yb=\pgf@y
    \pgfpathmoveto{\pgfpoint{\pgf@xa}{\pgf@ya}}
    \pgfpathlineto{\pgfpoint{\pgf@xb}{\pgf@ya}}
    \pgfpathmoveto{\pgfpoint{\pgf@xa}{\pgf@yb}}
    \pgfpathlineto{\pgfpoint{\pgf@xb}{\pgf@yb}}
 }
}
\makeatother

\usetikzlibrary{arrows}
\usetikzlibrary{calc}
\makeatletter

\usepackage{url}
\usepackage{amsmath}
\usepackage{amsfonts}
\usepackage{amssymb}
\usepackage{todonotes}
\definecolor{sangre}{rgb}{0.6,0.18,0.19}
\definecolor{dullmagenta}{rgb}{0.4,0,0.4}
\definecolor{darkblue}{rgb}{0,0,0.6}

\usepackage{relsize,etoolbox}
\usepackage[para,online,flushleft]{threeparttable}

\usepackage{amsmath}

\usepackage{bbm}

\usepackage{color, colortbl}
\definecolor{Gray}{gray}{0.95}
\usepackage{booktabs}
\usepackage{pdflscape}
\usepackage{amsfonts}
\usepackage{graphicx}
\usepackage{multirow}
\usepackage{epstopdf}
\usepackage{adjustbox}
\usepackage[titletoc,title]{appendix}

\definecolor{lavander}{cmyk}{0,0.48,0,0}
\definecolor{violet}{cmyk}{0.79,0.88,0,0}
\definecolor{burntorange}{cmyk}{0,0.52,1,0}

\def\lav{lavander!90}
\def\oran{orange!30}

\tikzstyle{peers}=[draw,circle,violet,bottom color=\lav,
                  top color= white, text=violet,minimum width=10pt]
\tikzstyle{superpeers}=[draw,circle,burntorange, left color=\oran,
                       text=violet,minimum width=20pt]
\tikzstyle{legendsp}=[rectangle, draw, burntorange, rounded corners,
                     thin,bottom color=\oran, top color=white,
                     text=burntorange, minimum width=2.5cm]
\tikzstyle{legendp}=[rectangle, draw, violet, rounded corners, thin,
                     bottom color=\lav, top color= white,
                     text= violet, minimum width= 2.5cm]
\tikzstyle{legend_general}=[rectangle, rounded corners, thin,
                           burntorange, fill= white, draw, text=violet,
                           minimum width=2.5cm, minimum height=0.8cm]

\usepackage[most]{tcolorbox}
 \tcbset{colback=white}

\usepackage{natbib}
\usepackage[margin=1in]{geometry}

\ifx\pdfoutput\undefined
\usepackage[hypertex,colorlinks,urlcolor = darkblue,citecolor = sangre,linkcolor=blue]{hyperref}
\else
\usepackage[pdftex,hypertexnames=false,colorlinks,urlcolor = darkblue,citecolor =  sangre,linkcolor=blue]{hyperref}
\fi

\makeatletter
\newcommand*\bigcdot{\mathpalette\bigcdot@{.5}}
\newcommand*\bigcdot@[2]{\mathbin{\vcenter{\hbox{\scalebox{#2}{$\m@th#1\bullet$}}}}}
\newcommand\EightPtClose{\@setfontsize\EightPtClose\@viiipt{9}}
\newcommand\TenPtType{\@setfontsize\TenPtType\@xpt\@xiipt}
\def\notesize{\TenPtType}
\def\notesize{\EightPtClose}
\newenvironment{figurenotes}[1][Note]{\begin{minipage}[t]{\linewidth}\notesize{\itshape#1: }}{\end{minipage}}
\makeatother

\begin{document}
\title{Common Subcontracting and Airline Prices\thanks{
We acknowledge the Bankard Fund for Political Economy at the University of Virginia for support. We thank Ha Pham for outstanding research assistance. We also thank participants and discussants at the 2022 SoAR Symposium, SEA 2022, IIOC 2023 and ABA Antitrust Law Section, for their suggestions.}}
\author{Gaurab Aryal\thanks{ Department of Economics, Washington University in St. Louis, \href{mailto:
aryalg@wustl.edu}{ aryalg@wustl.edu}} 
\and Dennis J. Campbell\thanks{ Department of Economics, University of Virginia, \href{mailto:
djc5yf@virginia.edu}{ djc5yf@virginia.edu}} 
\and Federico Ciliberto\thanks{ Department of Economics, University of Virginia, DIW and CEPR, \href{mailto:
ciliberto@virginia.edu}{ ciliberto@virginia.edu}} 
\and Ekaterina A. Khmelnitskaya\thanks{Sauder School of Business, University of British Columbia, \href{mailto:
ekaterina.khmelnitskaya@sauder.ubc.ca}{ekaterina.khmelnitskaya@sauder.ubc.ca}} 
}

\date{\today}
\maketitle

\begin{abstract}
In the US airline industry, independent regional airlines fly passengers on behalf of several national airlines across different markets, giving rise to \textit{common subcontracting}. On the one hand, we find that subcontracting is associated with lower prices, consistent with the notion that regional airlines tend to fly passengers at lower costs than major airlines. On the other hand, we find that \textit{common} subcontracting is associated with higher prices. These two countervailing effects suggest that the growth of regional airlines can have anticompetitive implications for the industry. 
\end{abstract}


\section{Introduction}

\begin{figure}[ht!]
\begin{center}
\caption{\bf Use of Regional Airlines \label{fig:regionalusage}}
 \includegraphics[scale=0.34]{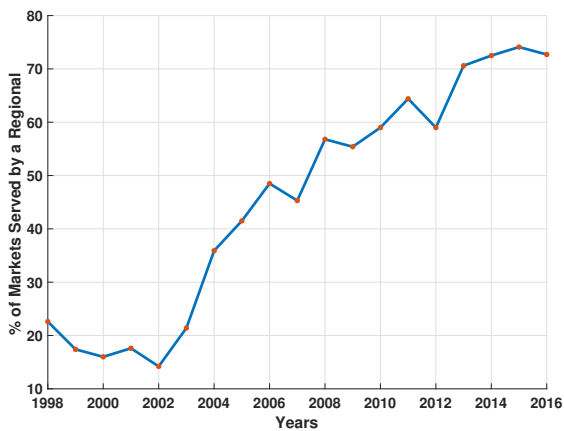}
\begin{figurenotes} The percentage of markets in our sample where operations of \textit{at least} one flight segment were subcontracted to an \textit{independent} regional airline. The observations are at the airline-market-(second quarter of a) year level. At its lowest, in 2002, regional airlines served at least one segment in 14.2\% of the (nonstop or connecting) markets. By 2016, use had risen to {72.7\%}
\end{figurenotes}
\end{center}
 \end{figure}

Over the last 20 years, national air carriers, e.g., American or Delta Airlines, have increasingly subcontracted their flight operations to regional airlines such as SkyWest and Trans State Airlines.\footnote{We use the terms ``national," ``legacy," and  ``major" airlines interchangeably to mean carriers that sell tickets. They include Alaska, American, America West, Continental, Delta, JetBlue, Northwest, Southwest, TWA, United, and US Air. Regional airlines do not sell tickets but only operate flights for major airlines.}  Figure \ref{fig:regionalusage} shows the dramatic growth in the use of regional carriers and underscores the changing nature of vertical structure in the airline industry. 

While regional carriers operate flights on behalf of only one national airline in a nonstop flight segment, they can serve several national airlines in different segments. For instance, in 1998, Trans States Airlines operated flights on behalf of Alaska, Delta, Northwest, Trans World, and United Airlines. We refer to this industry feature as \textit{common subcontracting}.  

We aim to determine whether this common subcontracting is associated with higher prices. To test our hypothesis, we use panel regressions of prices on a novel measure of common subcontracting that depends on the share of passengers transported through regionals in a given market \emph{and} the extent of overlapping use of a regional airline across markets by different national airlines. 

We use the US Department of Transportation \emph{Airline Origin and Destination Survey (DB1B)} dataset from 1998 to 2016 and data from the \emph{Official Aviation Guide of the Airways} (OAG) company to identify ticketing carriers and operating carriers. We supplement these data with ownership information from SEC 10-K filings and \emph{The Regional Airline Association} to identify independent regional airlines, like  SkyWest, from subsidiaries, like Envoy Air, a subsidiary of American. Henceforth, regional airline means an \emph{independent} regional airline. 

Our measure of common subcontracting is a function of the shares of passengers transported by competitors through regional carriers, which is likely to be endogenous in our panel regressions, even after including carrier, market, and year fixed effects. 
To address this concern, we construct instrumental variables based on \cite{ForbesLederman2009}. 

For each major airline, we use a measure of extreme weather conditions that the airline's \emph{competitors} face along the routes they use to serve a market. The identification assumption is that extreme weather conditions facing an airline's competitors are correlated with common subcontracting, but they do not directly affect the airline's prices. Additionally, we use the size of \textit{competitors'} networks of regional airlines as additional instruments as they are positively correlated with overlapping use of regional carriers and hence common subcontracting, but they should not directly affect prices.

To isolate the effect of common subcontracting on prices, we control for factor associated with possible efficiency gains from using smaller planes flown by regional carriers and competition among national airlines across multiple markets. Specifically, we control for the {share of passengers} transported by regional carriers in a market \citep{ForbesLederman2007, ForbesLederman2009, ForbesLederman2010, HeKosmopoulou2021, GilKimZanarone2022} and  multimarket contact among national airlines \citep{Edwards1955, BernheimWhinston1990, EvansKessides1994, CilibertoWilliams2014}, respectively.
These two control variables are also likely endogenous. We, use functions of the weather conditions as an IV for airline's decision to use regional airlines and its competitors' networks as an IV for multimarket contact. Section \ref{section:identification} discusses our identification strategy in detail.

We find that common subcontracting is associated with higher prices. In particular, an interquartile range (IQR) increase in common subcontracting increases prices by 1.8\%. 
Furthermore, this price effect of common subcontracting has increased in recent years. 
In particular, an IQR increase in common subcontracting pre-2004 did not affect prices because common subcontracting was negligible. However, in 2004-2012 and 2012-2016, an IQR increase in {common subcontracting} increased prices by $1.7\%$ and $5.1\%$, respectively. 

We also find that regional share is associated with \textit{lower} prices, which confirms that relying on regional airlines in a market should lower costs and, therefore, prices. Lastly, we confirm findings by \cite{EvansKessides1994, CilibertoWilliams2014} that multimarket contact increases prices. 

The price effects of common subcontracting suggest that the use of regional airlines \emph{per se} is beneficial, but their common use by national carriers--common subcontracting--is anticompetitive. Thus, we provide a novel insight into the vertical structure of the airline industry.
There is a vast literature on the role of the vertical structure of industry on market outcomes through foreclosure \citep{Coase1937, AlchianDemsetz1972, Williamson1979, NockeWhite2007}. Here, we have identified a new channel, common subcontracting, which is the exact \textit{opposite} of foreclosure. In this regard, our paper is similar to \cite{HortacsuSyverson2007}, who identify a non-foreclosure channel to exert market power.

Much remains to be learned of the underlying mechanism that rationalizes our results, but they are consistent with the conjecture that using common vertical relationships facilitates collusion, similar to the common ad agency in  \cite{BernheimWhinston1985} and the trade association in \cite{AwayaKrishna2020} that facilitate collusion.\footnote{Regional carriers employ pilots and stewards at the same wages and fly the same types of aircraft across different markets, independently of the major they serve, they likely reduce cost differences, increase co-movement of their costs, and lessen cost uncertainties across national airlines that use the same regional.}$^,$\footnote{Also see \cite{Vives1990, PiccoloThal2012} and \cite{ClarkHorstmannHoude2021}.} 

The paper is organized as follows. In Section \ref{section:regional}, we introduce regional airlines and subcontracting, and in Section \ref{section:CSCMMC}, define common subcontracting. In Sections \ref{section:result}- \ref{section:robustness}, we present our results. In Appendices \ref{section:sample}-\ref{section:weather}, we detail data constructions and the ``Weather IVs." 

\section{Regional Subcontracting\label{section:regional}}
Major airlines, or legacy airlines, operate large networks of flights across the US. 
Examples of major airlines are American Airlines (``AA"), Continental (``CO"), Delta (``DL"), United (``UA"), and US Airways (``US"). 
These airlines earn revenue by selling flight tickets within their network and, occasionally and primarily in the earlier years in our dataset, to flights in other major carriers' networks under code-share agreements. They set prices, schedule flights, manage seat inventories, and market tickets to passengers. 

Major airlines compete in many different markets, where a \emph{market} is defined as a unidirectional trip between any two airports.  
Airlines can serve a market via several combinations of flight segments, i.e., \textit{segment paths}, where a \textit{segment} is a unidirectional \textit{nonstop} leg between two airports. For example, in 2016, AA served the CHO-MCO (Charlottesville-Orlando) market via two \textit{segment paths}: CHO-CLT-MCO (through Charlotte) and CHO-PHL-MCO (through Philadelphia). The first \textit{segment path} consisted of two nonstop flight segments, one from CHO to CLT and the other from CLT to MCO. 

To transport passengers, major airlines either operate their flights, subcontract operations to regional airlines, or both. 
In the remainder of this section, we discuss some characteristics of this subcontracting relationship in the airline industry over time.

\subsection{Characteristics of the Subcontracting Relationship}

We have collected relevant information from major airlines and publicly owned regional airlines' 10-K filings with the SEC. In these 10-K filings, carriers report information regarding their contracts with each other, including the length of agreements, the scope of agreements, and, in certain instances, the payments. 

We observe subcontracting agreements (e.g., fixed-fee or revenue-sharing) between major and (independent) regional airlines. We also observe the start and end dates (if listed) for such agreements, the presence of termination clauses, and their terms. We collect some of these variables when available. Most regional airlines are publicly traded and work for multiple major airlines. In practice, subcontracting relationships are exclusive at the segment level, and a regional carrier serves only one legacy airline on a nonstop flight between two airports. However, a regional carrier can serve multiple legacy carriers in different routes. For example, SkyWest can serve AA in the CHO-PHL route and DL in the CHO-CLT route.

We also observe that the relationships between regional and major carriers span many years. On average, contracts between majors and independent regionals last 10.3 years, with some lasting as long as 24 years. 
Some contracts contain clauses for early termination.\footnote{For example, in its 2010 10-K filing,  Republic Airways Holdings notes:
\begin{quote} \emph{Delta may terminate the code-share agreements at any time, with or without cause, if it provides us 180 days' written notice for the E145 regional jet code-share agreement or after July 2015 for the E175 regional jet code-share agreement. For the E145 agreement, if Delta chooses to terminate any aircraft early, it may not reduce the number of aircraft in service to less than 12 during the 12 months following the 180-day initial notice period unless it completely terminates the code-share agreement.}\end{quote} } 

Lastly, most subcontracting agreements are ``fixed-fee" or ``capacity-purchase" agreements wherein a major airline sells tickets, often covers certain costs incurred during operations, and pays a fixed fee for service. Airlines may structure their capacity purchase contracts differently. For example, a major airline may supply fuel or lease aircraft to a regional carrier as part of a capacity purchase contract. Some contracts are ``revenue-sharing'' or ``pro-rate," where the regional airline receives a percentage of ticket revenue. However, such arrangements have become less common in recent years.

Although our data is incomplete, we find that the payments involved in these relationships can be large. On average, across all the years in our sample, major airlines paid \$1.7 billion per year to each regional carrier they contracted with, with some payments being as high as \$5.66 billion. The total payments have risen markedly over time, consistent with the increasing use of regional airlines (Figure \ref{fig:regionalusagetrend}) and the move towards a fixed-fee system. 
 
\cite{ForbesLederman2007, ForbesLederman2009} show that the availability of aircraft technologies determines whether a major airline contracts its service on a segment to a regional carrier. From the mid-to-late 1990s until the mid-2000s, airline manufacturers introduced new regional jets that allowed longer flights to be operated by small regional aircraft. Before these new regional jets, regional airlines typically operated turboprop planes with limited flight ranges. New regional jets have allowed major airlines to subcontract more flights to regional airlines as available regional jet technologies can be configured to comply with the major airline's needs.
For example, Embraer's E175SC jet is a special configuration of the E175 limited to 70 seats to take advantage of the performance improvements of the E175 while restricting the number of seats to 70.

\subsection{Ownership of Regional Airlines}

While regional airlines can be either subsidiaries of major airlines or independently owned, most are independent. For example, in 2016, seven of the ten largest regional airlines were independently owned, accounting for $66\%$ of all passengers served by regional airlines.

A major airline can subcontract with an independent regional or a subsidiary, but a subsidiary regional airline cannot operate on behalf of other major airlines.
For instance, AA owns Envoy, Piedmont, and PSA Airlines, and these regionals fly exclusively for AA. However, AA also contracts with independent regionals. Some carriers, for example, UA, subcontract solely to independent regional airlines.

\thispagestyle{empty}
\begin{table}[th!]
\caption{\bf Ownership Timeline\label{fig:chronology}}
\vspace{-0.1in}\hspace{-0.78in}\scalebox{0.85}{\begin{tabular}{llllll}
\toprule
{\bf Carrier Name} &
 {\bf Code }&
{\bf  Parent Company (1)} &
 {\bf Time (1) }&
{\bf  Parent Company (2)} &
 {\bf Time (2)} \\
  \midrule
Cape Air &
  9K &
  Hyannis Air Service, Inc. &
  1998 - now &
   &
   \\
Nantucket Airlines &
  ACK &
  Hyannis Air Service, Inc. &
  1998 - now &
   &
   \\
Midwest Airlines$^1$ &
  YX &
  Midwest Air Group, Inc$^{10}$ &
  1998 - 2008 &
  TPG Capital &
  2008 - 2009 \\
Skyway Airlines &
  AL &
  Midwest Air Group, Inc$^{10}$ &
  1998 - 2008 &
   &
   \\
Compass Airlines &
  CP &
  Northwest Airlines &
  2006 - 2008 &
  Delta Air Lines, Inc. &
  2008 - 2010 \\
Atlantic S.E. Airlines, Inc.$^2$ &
  EV &
   &
  1998 - 1999 &
  Delta Air Lines, Inc. &
  1999 - 2005 \\
Freedom Airlines, Inc. &
  F8 &
  Mesa Air Group, Inc. &
  2002 - 2010 &
   &
   \\
GoJet Airlines LLC &
  G7 &
  Trans States Holdings, Inc. &
  2005 - now &
   &
   \\
Big Sky Airlines &
  GQ &
  Big Sky Transportation Co. &
  1998 - 2002 &
  MAIR Holdings Inc.$^{11}$ &
  2002 - 2008 \\
Envoy Air$^{3}$&
  MQ &
  AMR Corp. &
  1998 - 2013 &
  American Airlines Group, Inc. &
  2013 - now \\
Comair &
  OH &
  Comair Holdings &
  1998 - 1999 &
  Delta Air Lines, Inc. &
  1999 - 2012 \\
Executive Airlines Inc. &
  OW &
  AMR Corp. &
  1998 - 2013 &
   &
   \\
Horizon Air &
  QX &
  Alaska Air Group Inc. &
  1998 - now &
   &
   \\
Republic Airlines &
  YX$^5$ &
  Republic Airways Holdings &
  2004 - now &
   &
   \\
Mesaba Airlines &
  XJ &
  MAIR Holdings Inc.$^{11}$ &
  1998 - 2007 &
  Northwest Airlines Corp. &
  2007 - 2008 \\
Air Midwest &
  ZV &
  Mesa Air Group, Inc. &
  1998 - 2008 &
   &
   \\
Air Wisconsin Airlines &
  ZW &
  CJT Holdings &
  1998 - now &
   &
   \\
Endeavor Air$^4$&
  9E &
  Northwest Airlines &
  1998 - 2003 &
  Pinnacle Airlines Corp. &
  2003 - 2013 \\
Trans States Airlines &
  AX$^{6}$ &
  Trans States Holdings, Inc. &
  1998 - now &
   &
   \\
Colgan Air &
  9L &
   &
  1998 - 2007 &
  Pinnacle Airlines Corp. &
  2007 - 2012 \\
SkyWest Airlines &
  OO &
  SkyWest, Inc. &
  1998 - now &
   &
   \\
Chautauqua Airlines &
  RP &
  Wexford Capital$^{12}$ &
  1998 - 2004 &
  Republic Airways Holdings &
  2004 - 2014 \\
Shuttle America Corp. &
  S5 &
   &
  1998 - 2001 &
  Wexford Capital$^{12}$ &
  2001 - 2005 \\
ExpressJet Airlines, Inc. &
  XE$^{7}$ &
  Continental Airlines &
  1998 - 2002 &
  ExpressJet Holdings Inc. &
  2002 - 2010 \\
Mesa Airlines &
  YV &
  Mesa Air Group Inc. &
  1998 - 2011 &
  Mesa Air Group Inc. &
  2011 - now \\
America West Airlines &
  HP &
  America West Holdings Corp. &
  1998 - 2005 &
  US Airways Group Inc. &
  2005 - 2007 \\
Business Express Airlines &
  HQ &
   &
  1998 - 1999 &
  AMR Corp. &
  1999 - 2000 \\
Trans World Airlines &
  TW &
   &
  1998 - 2001 &
  AMR Corp. &
  2001 - 2001 \\
Frontier Airlines &
  F9 &
   &
  1998 - 2006 &
  Frontier Airlines Holdings, Inc &
  2006 - 2009 \\
Scenic Airlines &
  YR &
  SkyWest Inc. &
  1998 - 2007 &
  Grand Canyon Airlines$^{14}$ &
  2007 - 2009 \\
PSA Airlines &
  OH$^{8}$&
  US Airways &
  1998 - 2013 &
  American Airlines Group, Inc. &
  2013 - now \\
USAir Shuttle &
  TB &
  US Airways Group, Inc. &
  1998 - 2000 &
   &
   \\
UFS Inc. &
  U2 &
  Trans States Holdings, Inc. &
  1998 - 2000 &
   &
   \\
Lynx Aviation &
  L3$^9$ &
  Frontier Airlines Holdings, Inc. &
  2006 - 2009 &
  Republic Airways Holdings &
  2009 - 2011\\

Piedmont Airlines &
    PT &
    US Airways &
    1998 - 2013 &
    American Airlines Group, Inc. &
    2013 - now\\

Allegheny Airlines &
    AL &
    US Airways &
    1998 - 2004$^{13}$&
     &
    \\
  \bottomrule
\end{tabular}}
\begin{figurenotes}
Ownership of regional airlines and the corresponding event years. This information is collected by the authors from various sources including the SEC 10K filings and airlines' webpages. For several airlines their names have changed and they are recorded with a numbered notes as follows: (1) Midwest Express Airlines before 2003; (2)  ExpressJet Airlines, Inc. since 2011; (3) American Eagle Airlines, Inc. until 2014;  (4) Express Airlines I before 2002 and Pinnacle Airlines, Inc before 2013; (5)  RW before 2009; (6) also used the code 9N; (7) also used the code RU; (8) used code 16 before 2013; (9) L4 before 2009, and 0IQ before 2007; (10) Midwest Express Holdings, Inc. before 2004; (11) Mesaba Holdings Inc. until August 2003;  (12)  Wexford Management, LLC before 2000; (13) merged into Piedmont Airlines; (14) absorbed by Grand Canyon Airlines; contd.
\end{figurenotes}
\end{table}
\thispagestyle{empty}
\begin{table}[ht!]
\caption*{Table 1: {\bf Ownership Timeline} (continued)}\vspace{-0.1in}
\hspace{-0.7in}\scalebox{0.85}{\begin{tabular}{llllll}
\toprule
{\bf Carrier Name} &
 {\bf Code} &
  {\bf Parent Company (3)} &
  {\bf Time (3)} &
  {\bf Parent Company (4)} &
  {\bf Time (4)} \\
  \midrule
Cape Air                  & 9K                                   &                       &            &  &  \\
Nantucket Airlines        & ACK                                  &                       &            &  &  \\
Midwest Airlines$^1$  &
  YX &
  Republic Airways Holdings &
  2009 - 2009 &
   &
   \\
Skyway Airlines           & AL                                   &                       &            &  &  \\
Compass Airlines          & CP                                   & Trans States Holdings & 2010 - now &  &  \\
Atlantic S.E. Airlines, Inc.$^2$ &
  EV &
  SkyWest, Inc. &
  2005 - now &
   &
   \\
Freedom Airlines, Inc.    & F8                                   &                       &            &  &  \\
GoJet Airlines LLC        & G7                                   &                       &            &  &  \\
Big Sky Airlines          & GQ                                   &                       &            &  &  \\
Envoy Air$^3$ &
  MQ &
   &
   &
   &
   \\
Comair                    & OH                                   &                       &            &  &  \\
Executive Airlines Inc.   & OW                                   &                       &            &  &  \\
Horizon Air               & QX                                   &                       &            &  &  \\
Republic Airlines         & YX$^5$                 &                       &            &  &  \\
Mesaba Airlines &
  XJ &
  Delta Air Lines, Inc. &
  2008 - 2010 &
  Pinnacle Airlines Corp. &
  2010 - 2012$^{17}$ \\
Air Midwest               & ZV                                   &                       &            &  &  \\
Air Wisconsin Airlines    & ZW                                   &                       &            &  &  \\
Endeavor Air$^4$ &
  9E &
  Delta Air Lines &
  2013 -now &
   &
   \\
Trans States Airlines     & AX$^6$                                   &                       &            &  &  \\
Colgan Air                & 9L                                   &                       &            &  &  \\
SkyWest Airlines          & OO                                   &                       &            &  &  \\
Chautauqua Airlines       & RP                                   &                       &            &  &  \\
Shuttle America Corp. &
  S5 &
  Republic Airways Holdings &
  2005 - 2017$^{15}$ &
   &
   \\
ExpressJet Airlines, Inc. &
  XE$^7$ &
  SkyWest Inc &
  2010 - 2011$^{16}$ &
   &
   \\
Mesa Airlines             & YV                                   &                       &            &  &  \\
America West Airlines     & HP                                   &                       &            &  &  \\
Business Express Airlines & HQ                                   &                       &            &  &  \\
Trans World Airlines      & TW                                   &                       &            &  &  \\
Frontier Airlines &
  F9 &
  Republic Airways Holdings &
  2009 - 2013 &
  Indigo Partners LLC &
  2013 - now \\
Scenic Airlines           & YR                                   &                       &            &  &  \\
PSA Airlines              & OH$^8$                  &                       &            &  &  \\
USAir Shuttle             & TB                                   &                       &            &  &  \\
UFS Inc.                  & U2                                   &                       &            &  &  \\
Lynx Aviation             & L3$^9$ &                       &            &  & \\

Piedmont Airlines         & PT &                       &            &  & \\

Allegheny Airlines         & AL &                       &            &  & \\
\bottomrule
\end{tabular}}
\begin{figurenotes}
(continued) (15)  merged into Republic Airlines); (16) merged into Atlantic Southeast Airlines, but code XE was still used till January 2012; and (17)  merged into Pinnacle Airlines. 
\end{figurenotes}
\end{table}

Table \ref{fig:chronology} presents a chronology of regional airline ownership information from 1998 to 2016 and is based on information we collected from various airlines' 10-K filings and by searching across airlines' websites. Each row corresponds to a regional carrier and its owner, and each event records the year the ownership changed. 
For example, consider Shuttle America (S5). This regional airline was independently owned at the start of our sample and was purchased by Wexford Capital in 2001. In 2005, it became a part of Republic Airways Holdings,  with regional airlines Chautauqua (RP) and Republic Airlines (YX). Republic Airways Holdings is not affiliated with any major airline, so Shuttle America is an independent regional airline. 

\subsection{Regional Airline Usage Patterns}

We use the US DOT's DB1B dataset to investigate regional airline usage and how it has changed over time. This dataset consists of a 10\% sample of domestic airline tickets. We use data from every second quarter from 1998 to 2016 and define a \textit{market} as a unidirectional trip between two airports, irrespective of intermediate transfer points.\footnote{See Appendix \ref{section:sample} for more detail on the sample construction.}

The DB1B database reports the \textit{ticketing} carrier for service in a market for each ticket sold in a quarter, as well as the \textit{operating} carrier(s) that transported the passenger along the \textit{segment path} used for each ticket. Essentially, we observe the \textit{segment path} for each ticket sold. We observe $16$ different major airlines, including legacy airlines mentioned above, low-cost carriers such as Southwest (WN) and JetBlue (B6), and $29$ regional carriers. Some regional carriers are subsidiaries of major airlines, e.g., Envoy and Horizon, and others are independently owned, e.g., Skywest, Mesa, and ExpressJet.\footnote{We confirm operating information for regional carriers using OAG Market Intelligence-Schedules dataset.} Our final sample has $3,499,463$ unique market-year-ticketing carrier-\textit{segment path} observations from $25,767$ unique markets.

Table \ref{tab:markets19982016} presents a snapshot of the regional airline landscape at the start, middle and end of our sample that includes the name of the regional airline, its IATA code, all major airlines it worked for, number of markets served, and its percentage of all unique markets.\footnote{The numbers in Table \ref{tab:markets19982016} may add up to more than 100\% because airlines may use different \textit{segment paths} to serve a market, and a route may be served by different regional airlines.}

Regional usage has been increasing over time. In 1998, American Eagle, a regional airline owned by AA, operated in 6,272 markets, or 30.8\% of the total 20,373 markets. The most-used independent regional carrier in 1998 was Mesaba Airlines, which operated in 4,296 markets or 21.1\% of the total. In 2016, the most-used regional airline was ExpressJet, an independent regional airline operating in 15,332 markets, 70.5\% of the 21,783 markets.

\begin{figure}[t!]
\caption{\bf Usage of Regional Airlines \label{fig:regionalusagetrend}}
\begin{center}
\includegraphics[scale=0.4]{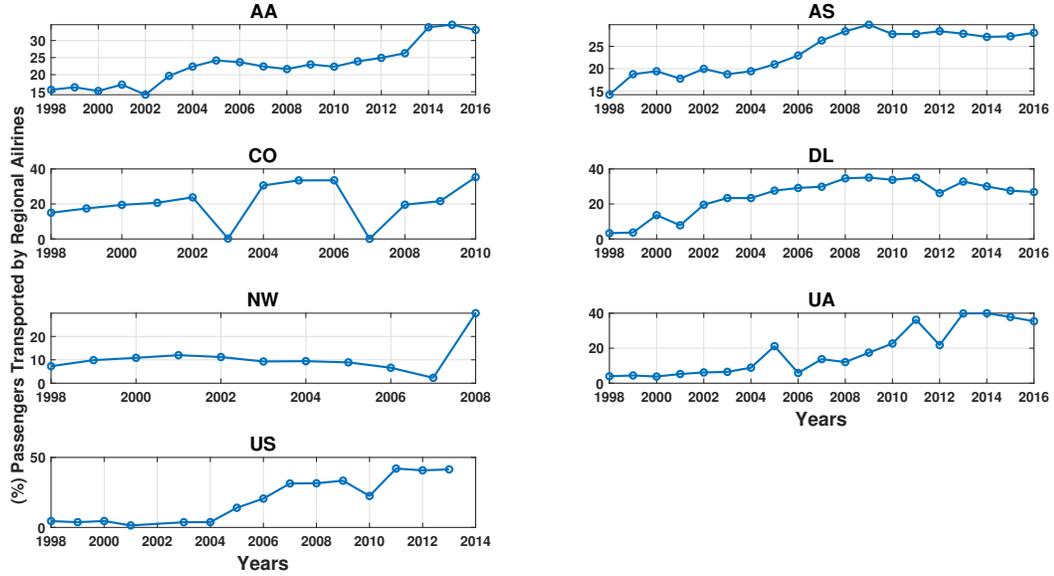}  
\begin{figurenotes} The time trends of the percentage of total passengers transported by regional airlines for each legacy airline.
\end{figurenotes}
\end{center}
 \end{figure}

Figure \ref{fig:regionalusagetrend} displays the share of passengers that (subsidiaries and independent) regional airlines transport on behalf of major airlines. It shows a striking rise in the usage of regionals over time. 
For instance, from 1998 to 2016, AA increased its percentage from 15.6\% to 33.1\%.

\subsection{Regional Share}
We consider the share of passenger-seat miles transported through (independent) regional airlines. We refer to this variable as ``regional share," and denote it by $\texttt{Regional Share}_{j,m,t}$, in year $t=1,\ldots, T$, market $m=1, \ldots, M$, and major airline $j\in n:= \left\{ \text{AA, AS, B6, CO}, \ldots\right\}$. The variable $\texttt{Regional Share}_{j,m,t}$ measures the reliance of a (national) airline on regional airlines in a particular market at a given time. We consider both the number of seats flown by the regional carrier and the miles to facilitate the interpretation of this variable as a proxy
\newpage
{\fontsize{11}{10}\selectfont {\begin{longtable}[t!]{lcllc}
\caption{\bf Markets Served by Regional Airlines} \label{tab:markets19982016} \\
\toprule \multicolumn{1}{c}{\textbf{Regional}} & \multicolumn{1}{c}{\textbf{Code}} & \multicolumn{1}{c}{\textbf{Airlines Served}}&
 \multicolumn{1}{c}{\textbf{\# of Mkts}} & \multicolumn{1}{c}{\textbf{\% of Total Mkts}} \\ 
 \hline
  {\bf 1998:} Trans States Airlines & 9N  & AS, DL, NW, TW, UA, US &          2,486 & 12.2 \\
  Atlantic Southeast & EV  & DL &          4,059 & 19.9 \\
  American Eagle & MQ  & AA (Owned) &          6,272 & 30.8\\
  Horizon Air & QX  & AS (Owned) &           712 & 3.5\\
  ExpressJet & RU  & CO (Owned) &          5,623 & 27.6\\
  US Air Shuttle & TB  & US (Owned)  &           166 & 0.8\\
  UFS & U2  & UA &           886 & 4.3\\
  Mesaba Airlines & XJ  & NW (Owned, until 2007)&          4,296 & 21.1\\
  Air Wisconsin & ZW  & UA &          1,267 & 6.2\\
 {\bf 2007:} PSA Airlines           & 16 & UA, US & 4,883 & 24.6\\
  ExpressJet & EV  & CO, DL, NW &          10,897 & 55.0\\
  GoJet Airlines & G7  & UA, US &          1,542 & 7.8\\
  American Eagle & MQ  & AA (owned) &          9,085 & 45.8\\
  Comair & OH  & DL (owned) &          6,433 & 32.5\\
  Executive Airlines & OW  & AA (owned) &           167 & 0.8\\
  Horizon Air & QX  & AS (owned) &           1,495 & 7.5\\
  Republic Airlines & RW  & F9, UA, US &          3,175 & 16.0\\
  Shuttle America & S5  & CO, DL, NW, UA, US &          3,214 & 16.2\\
  Mesaba Airlines & XJ  & CO, DL, NW &           1,843 & 9.3\\
  Mesa Airlines & YV  & UA, US &          7,385 & 37.3\\
  Air Wisconsin & ZW  & UA, US &          3,705 & 18.7\\
 {\bf 2016:} Endeavor Air & 9E  & DL (owned) &         7,361 & 33.9\\
  Compass Airlines & CP  & AA, AS, DL &          3,625 & 16.7\\
  ExpressJet & EV  & AA, AS, DL, UA &         15,332 & 70.5\\
  GoJet Airlines & G7  & DL, UA &          4,940 & 22.7\\
  Envoy Air & MQ  & AA (owned) &          9,234 & 42.5\\
  PSA Airlines & OH  & AA (owned) &         8,595 & 39.5\\
  Skywest & OO  & AA, AS, DL, UA &         12,333 & 56.7\\
  Horizon Air & QX  & AS (owned) &          1,150 & 5.3\\
  Shuttle America & S5  & DL, UA &          4,583 & 21.1\\
  Mesa Airlines & YV  & AA, AS, UA &          7,588 & 34.9\\
  Air Wisconsin & ZW  & AA &          4,315 & 19.9\\
\bottomrule
\end{longtable}}}
\begin{figurenotes} Rows correspond to regional airlines. The first column is its name; the second is its two-digit code. The third column is the set of all major airlines the regional airline has worked for that year. The fourth column is the total number of markets the regional airline serves. The fifth column is the percentage of regional airlines' total markets.
\end{figurenotes}
\vspace{1em}

\noindent for the utilization cost associated with subcontracting. 
We consider both the number of seats flown by the regional carrier and the miles to facilitate the interpretation of this variable as a proxy for the utilization cost associated with subcontracting. This measure is important for our empirical analysis because it is a proxy for costs, and we expect it to lower prices.

For an illustration of how we calculate the share, consider an example in Table \ref{table:exampleofcscpart1}. 
 Suppose we consider the Charlottesville to Dallas Fort Worth (CHO-DFW) market being served by AA and DL. Suppose AA serves the market through either Charlotte, i.e., CHO-CLT-DFW, or Philadelphia, i.e., CHO-PHL-DFW, and that 100 passengers fly via CLT and 50 via PHL. Now, suppose that OO serves the CHO-CLT segment on behalf of AA, and no other regionals are used in this market.

\begin{table}[t!]
\caption{\bf{Example of Market-\textit{Segment Paths} Information}}\label{table:exampleofcscpart1}
\hspace{-0.6in}\begin{tabular}{ c c c c c c c c c }
\toprule
{\bf Major} &{\bf Mkt} &{\bf Seg1} &{\bf Carrier1} &{\bf Seg1 Dist} &{\bf Seg2} & {\bf Carrier2} &{\bf Seg2 Dist} & {\bf\#Pass} \\
\midrule
AA & CHO-DFW & CHO-CLT & OO & 150 & CLT-DFW & AA & 300 & 100 \\
AA & CHO-DFW & CHO-PHL & AA & 50 & PHL-DFW & AA & 450 & 50 \\
DL & CHO-DFW & CHO-ATL & DL & 200 & ATL-DFW & DL & 250 & 200 \\
\bottomrule
\end{tabular}
\begin{figurenotes}
Example of usage of regional airlines. The first column of the table, ``Major,'' lists the airline that sold tickets for the given \textit{segment path}. The second column, ``Market,'' lists the origin and destination airport of the market. The column ``Seg1'' lists the first flight segment along the given \textit{segment path}, ``Carrier1'' lists the carrier's code which operated that flight, and ``Seg1 Dist'' lists the nonstop distance of flight segment 1. The last column, ``\#Pass,'' is the number of passengers transported along the given \textit{segment path} and who purchased a ticket from the major carrier.  \end{figurenotes}
\end{table}

Here, AA uses OO to transport 100 passengers in the CHO-CLT segment, which is 150 (nautical) miles. 
 The total passenger seat-miles transported by AA is $100\times (150+300) + 50 \times (50+450)=70,000$, out of which $100\times 150=15,000$ miles are through OO. 
 Therefore, the regional share for AA in the CHO-DFW market is 0.21.
 Likewise, the regional share for DL in the CHO-DFW market is 0.
 
\begin{figure}[ht!]\caption{\bf Regional Share \label{fig:reg_share_average}}
\begin{center}    \includegraphics[scale=0.43]{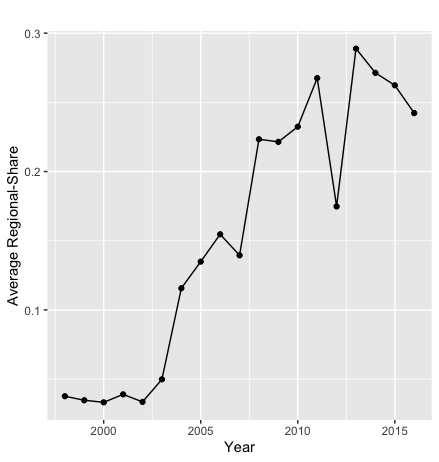}
 \begin{figurenotes} Yearly average of regional-share, where the average is taken across all national airlines and markets. 
 \end{figurenotes}\end{center}
 \end{figure}

  We calculate such shares for each national carrier, year, and market. The average regional share across our sample is $0.154$, with a median of $0$. In Figure \ref{fig:reg_share_average}, we present the yearly average of regional-share, where the average is taken across all airlines and all markets. We observe that regional share has increased over time. 

\section{Common Subcontracting and Multimarket Contact \label{section:CSCMMC}}

\subsection{Common Subcontracting}\label{section:definecsc}
We now discuss the overlapping use of independent regional airlines by major airlines and present our novel measure, or index, of the extent of this phenomenon, which we refer to as \emph{common subcontracting}. 
 
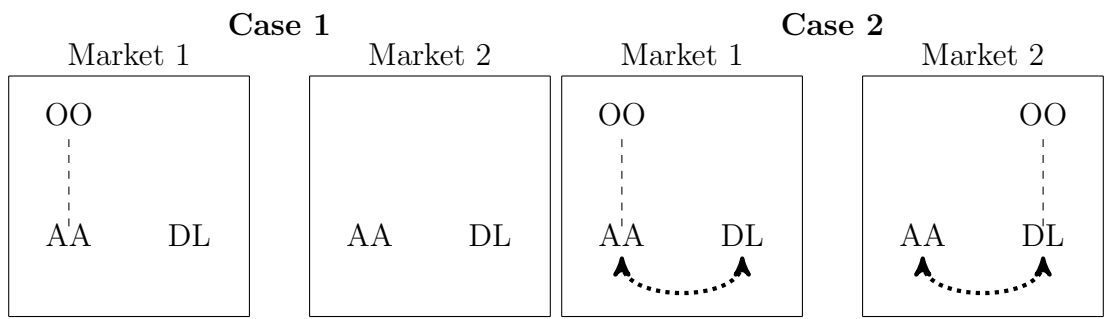
\begin{figure}[ht!]
\begin{center}
\caption{\bf Example of Common Subcontracting \label{fig:cscexample}}
\vspace{0.4em}
\begin{tikzpicture}[scale=0.8]
\node[above] at (4.5, 4.5) {{\bf Case 1}};
\draw (0,0) -- (0,4);
\draw (0,4) -- (4,4);
\draw (4,4) -- (4,0);
\draw (4,0) -- (0,0);
\node[above] at (2,4) {Market 1};
\node[above] at (1,1) {AA};
\node[above] at (3,1) {DL};
\node[above] at (1,3) {OO};
\draw[dashed] (1,1.5)--(1,3);
\draw (5,0) -- (5,4); \draw (5,4) -- (9,4); \draw (9,4) -- (9,0); \draw (9,0) -- (5,0);
\node[above] at (7,4) {Market 2};
\node[above] at (6,1) {AA}; 
\node[above] at (8,1) {DL};
\end{tikzpicture}
\begin{tikzpicture}[scale=0.8]
\vspace{1em}
\node[above] at (4.5, 4.5) {\bf Case 2};
\draw (0,0) -- (0,4);
\draw (0,4) -- (4,4);
\draw (4,4) -- (4,0);
\draw (4,0) -- (0,0);
\node[above] at (2,4) {Market 1};
\node[above] at (1,1) {AA};
\node[above] at (3,1) {DL};
\node[above] at (1,3) {OO};
\draw[dashed] (1,1.5)--(1,3);
\draw (5,0) -- (5,4); \draw (5,4) -- (9,4); \draw (9,4) -- (9,0); \draw (9,0) -- (5,0);
\node[above] at (7,4) {Market 2};
\node[above] at (6,1) {AA}; 
\draw[dashed] (8,1.5)--(8,3);
\node[above] at (8,1) {DL}; \node[above] at (8,3) {OO};
\draw [dotted, ultra thick, <->] (1,1) to [out=270, in=270]  (3,1);
\draw [dotted, ultra thick, <->] (6,1) to [out=270,in=270] (8,1);
\end{tikzpicture}
\begin{figurenotes}
Example of overlapping use of regional airlines. Case 1: only AA uses a regional airline (OO) in Market 1 (denoted by the dotted line connecting OO and AA), so there is no overlap. Case 2: AA uses OO in Market 1, and DL uses OO in Market 2, creating an overlapping use of regional between AA and DL (denoted by the bold dashed lines between these two airlines).   
\end{figurenotes}\end{center}
\end{figure}

We use a stylized example to illustrate how we construct our measure. 
Suppose two major airlines (AA and DL) operate in two markets (1 and 2). 
First, suppose only AA subcontracts its operations to a regional airline (OO) in Market 1. DL does not use any regional in either of the two markets. Here, there is no overlapping use of regional airlines. Now suppose that DL subcontracts its Market 2 operations to OO, resulting in an overlapping use of OO by AA and DL. These two cases are shown in Figure \ref{fig:cscexample}. 

We take the following steps to construct the index of common subcontracting.
First, we take all the \textit{segment paths} each major airline uses to serve a market and record whether the major airline subcontracts operations of one or more flight segments to regional airlines. 

Second, we calculate the percentage of passengers operated by each independent regional carrier on behalf of each major airline at the \textit{market} level across all possible \textit{segment paths}. Consider the example in Table \ref{table:exampleofcscpart1} again. Here the percentage of passengers served by carrier OO in this market on behalf of AA is $\frac{100}{150}=0.667$.  

Third, we use data across all markets to determine for each period $t$ and each major airline $j$ the names of all regional airlines it subcontracts, thereby determining which major and regional airlines have vertical relationships somewhere within the larger network of flights. 

Fourth, we use the passenger percentages served by regional carriers and information on other major airlines that subcontract with these regionals \textit{anywhere} in the network to determine the overlap between each major airline and each of its \emph{competitors in the market}. 
Returning to Table \ref{table:exampleofcscpart1}, recall that only AA subcontracts out to regional airlines, and it subcontracts 66.7\% of its passengers to regional carrier OO. If DL has a relationship with OO from other flight segments, we say that DL overlaps with 66.7\% of AA's operations in this market. By contrast, AA has 0\% overlap with DL's operations since no other regionals are involved in that market.\footnote{Here, we have two national airlines. So there are two unidirectional carrier-pair-specific measures. In markets with three major airlines, we obtain six different measures.} 

Fifth, and finally, we average the information shares calculated in step four across all unidirectional major carrier pairs in a given market to measure the average common subcontracting in each market. In the example market with AA and DL, we have an average measure of common subcontracting equal to 33.3\%.\footnote{Suppose DL subcontracts with OO to transport all its 200 passengers in Segment 1. 
Then, AA would have an overlap of 100\%, and the common subcontracting would be $(100\%+66.7\%)/2=83.5\%$.} Note that in Case 1 in Figure \ref{fig:cscexample}, common subcontracting is zero because there is no overlapping usage of a regional airline.  

Next, we define the measure formally. Let $i$ and $j$ denote major airlines where $i,j\in n:= \left\{ \text{AA, AS, B6, CO}, \ldots\right\}$. Following the steps described above, we define the level of \textit{common subcontracting} in market $m$ in period $t$ as 
\begin{eqnarray}
\texttt{CSC}_{mt}=\frac{1}{n\left( n-1\right) |K_m|}\sum_{i=1}^{n}\sum_{j\neq
i}\sum_{k\in K_m}s_{kjmt}\times B_{ikt},\label{eq:csc}
\end{eqnarray}
where $s_{kjmt}$ is the share of passengers in $m$ transported by the regional carrier $k\in K_m$ on behalf of major carrier $j$ in market $m$ at time $t$; $B_{ikt}=1$ if carrier $i$ subcontracts with regional carrier $k$ in \emph{at least one (potentially different) market}, and $B_{ikt}=0$ otherwise. 
So, for example, if DL subcontracts OO in market $m^{\prime }\neq m$, then $B_{DL, OO, mt}=1.$
Thus, $\texttt{CSC}$ is an average across major carriers of the share of passengers transported by shared regional airlines. 

The common subcontracting index takes values between 0 and 1 and is defined at the market level. A zero value means that none of the shared regional airlines transport passengers. In contrast, common subcontracting of one means that all of the passengers in the market are transported by shared regional airlines.

\begin{table}[t!]
\centering
\caption{\bf{Summary Statistics}}
\begin{tabular}{lcccc} \toprule
 & Mean & Median & St. Dev & \\ \midrule
{\bf Market-Specific} &  &    &  &  \\ 
Common Subcontracting & 0.161 & 0.045 & 0.242 &  \\
Multimarket Contact  & 5.559 & 5.048 & 2.689 &  \\
{\bf Carrier-Market-Specific} &    &  &  &  \\
Regional Share& 0.154& 0 &0.274\\ 
Price (\$)  & 258.14 & 242.56 & 95.33 \\ 
Log Price (\$) & 5.495&5.491&0.339\\
Network- Origin (in 1,000) & 0.843 & 0.870 & 0.299 \\
Network- Destination (in 1,000) & 0.841 & 0.870 & 0.298 \\
 \bottomrule
\end{tabular}
\begin{figurenotes}
Summary statistics of key variables based on a sample of 1,001,835 carrier-market-time observations across 279,958 market-time observations. Common subcontracting is defined in Equation (\ref{eq:csc}), and multimarket contact is defined in Equation (\ref{eq:ekmmc}) (in 1,000). Price refers to one-way fare expressed in 2012 US dollars; regional share is the share of passenger seat miles transported through regionals and is defined at the major carrier and market level;  the network is the number of markets served out of the origin-destination airports (in 100).
\end{figurenotes}
\label{table:sumstats}
\end{table} 

Table \ref{table:sumstats} shows that the average level of $\texttt{CSC}_{mt}$ in our sample across market-time observations is $0.161$,  while the median is $0.045$. The low relative median value is because there was relatively little regional usage in the early years of our sample (see Figure \ref{fig:regionalusagetrend}).

\begin{figure}[ht!]
\caption{\bf Common Subcontracting and Multimarket Contact \label{fig:csc_mmc}}
 \begin{center}
 \includegraphics[scale=0.45]{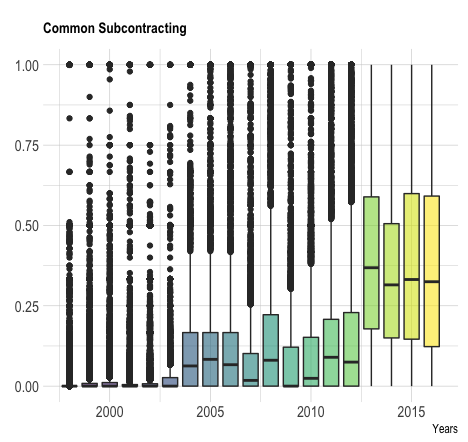}~
  \includegraphics[scale=0.45]{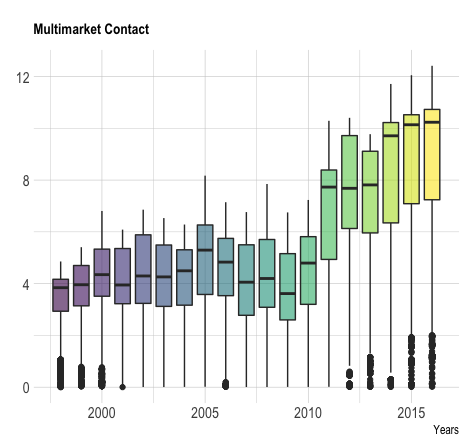}
 \begin{figurenotes} Box plots (with whiskers) of \textit{common subcontracting} across markets (left) and multimarket contact (right) by year. 
 \end{figurenotes}
  \end{center}
 \end{figure}

Figure \ref{fig:csc_mmc} (left panel) displays box plots of common subcontracting from 1998 until 2016. We observe that over our sample period, the level of the common subcontracting present in markets increased substantially. For instance, the 50\textsuperscript{th} and 75\textsuperscript{th} percentiles have gone from 0 in 2003 to 0.18 and 0.5, respectively, in 2016. In that time, the unweighted mean of common subcontracting across markets increased by 1,350\% from 0.02 to 0.29.
The figure shows a modest increase in the 2005 followed by a  rapid increase in 2012 that
coincided with bankruptcies \citep{CilibertoSchenone2012}, consolidation among regional airlines, 
and with the introduction of new regional jet technologies. These mergers were between DL and NW, UA and CO, and US and AA.

\subsection{Multimarket Contact}

Next, we define multimarket contact following the existing literature. See  \cite{EvansKessides1994} and \cite{CilibertoWilliams2014} and the references therein. We denote this measure of contact by {\texttt{MMC}}, and it is the average of the total pairwise contacts of the firms in a market. 

To define this measure more formally, we introduce the following notation. 
Let $i$ and $j$ be defined as before and denote major airlines where $i,j\in n:= \left\{ \text{AA, AS, B6, CO}, \ldots\right\}$. 
For an airline pair $\{i,j\}\in n \times n $ and a market $m\in M$, let $D_{ijmt}\in\{0,1\}$ be a dummy variable equal to one if both $i$ and $j$ are present in market $m$ in time $t$ and zero otherwise. Let $r_{ijt}:=\sum_{m\in M} D_{ijmt}$ be the total number of pairwise contacts over all the markets served by both $i$ and $j$ in period $t$.
Then, as \cite{EvansKessides1994}, multimarket contact is  
\begin{eqnarray}
\texttt{MMC}_{mt} = \frac{{\m}_{mt}}{N_{\m_{mt}}}:=\frac{\sum_{i=1}^n \sum_{\substack{j=1 \\ j\neq i}}^n \left(r_{ijt}\times D_{ijmt}\right)}{{\sum_{i=1}^n \sum_{\substack{j=1 \\ j\neq i}}^n D_{ijmt}}},\label{eq:ekmmc}
\end{eqnarray}
where ${\m}_{mt}$ is the total pairwise contact in market $m$  and $N_{\m_{mt}}$ is the total number of possible pairwise contacts, divided by 1,000. 

Table \ref{table:sumstats} shows that the average of multimarket contact is 5.559. This value is smaller than the average values in \cite{EvansKessides1994} and \cite{CilibertoWilliams2014} because our sample includes all MSAs to ensure that we also capture smaller markets.

Figure \ref{fig:csc_mmc} (right panel) displays box plots for multimarket contact across markets from 1998 until 2016.
Median multimarket contact more than doubled from 3.93 in 1998 to 7.65 in 2016, and the 75th percentile increased by 170\%, from 3.97 to 10.73. Of particular note are the jumps from 2010 to 2011 and 2013 to 2014. 
The unweighted mean of multimarket contact across all markets almost doubled from 4.26 pre-2010 to 7.96 post-2010. 
As with common subcontracting, these changes in multimarket contact coincide with mergers of major airlines. 

 \section{Empirical Analysis\label{section:result}}
We describe the ticket prices and carriers' networks at the origin and destination airports. Then, we introduce the empirical specification and discuss identification and results. 

\subsection{Prices and Network Service}
 Average fares are calculated at the carrier-market-time level, deflated using the 2012 consumer price index. 
 Whenever a ticketing carrier operates more than one \textit{segment path} to serve a market, we use weighted average prices from all \textit{segment paths} where the weights are the share of passengers transported along each \textit{segment path}. In other words, the weighted average price is the ratio of total revenue in a market to the total number of passengers.
The average fare in the sample is $\$258.14$ with a standard deviation of $\$95.33$. See Table \ref{table:sumstats}.
 
Next, we introduce $\texttt{Network Origin}_{jmt}$ and $\texttt{Network Destination}_{jmt}$ as additional controls that measure the airline's ability to offer differentiated products at a given market and time \citep{Berry1990, Berry1992, CilibertoTamer2009}. These variables are the number of markets (in 100s) airline $j$ serves from the origin and destination airports of market $m$ in time $t$, irrespective of any stops. For example, suppose AA serves the CHO-DFW market via CLT. Then the origin and destination networks are the numbers of airports AA serves from CHO and DFW, respectively. 

We contend that these measures of an airline's network are exogenous relative to prices in any particular year and market because setting up an airline network is a sticky process that takes many years and requires sunk start-up investments. As shown in Table \ref{table:sumstats}, airlines serve an average of 84 markets from an airport. 

\subsection{Empirical Specification}
We hypothesize that, on average, prices are higher in markets with either high common subcontracting after accounting for the lower cost associated with using regional carriers by airlines and higher prices from multimarket contact.
To evaluate this hypothesis, we use panel data and estimate a two-way fixed-effects model:
\begin{eqnarray}
		\ln(\texttt{price}_{j,m,t})&=&\beta_0 + \gamma_{j} + \gamma_{m} + \gamma_{t} +  \texttt{CSC}_{m,t}\beta_{csc}+ \texttt{Regional Share}_{j,m,t}\beta_{share}+{\bf \texttt{MMC}}_{m,t}\beta_{mmc}\notag\\&&\!\!\!+ \texttt{Network Origin}_{j,m,t}\gamma_{origin}+ \texttt{Network Destination}_{j,m,t}\gamma_{dest}+ \varepsilon_{j,m,t},\quad\label{eq:model_mmccsc}
\end{eqnarray}
 where the dependent variable is the log of the ticket (one-way) price sold by major airline $j$ in market $m$ in year $t$, and $\gamma_{j}, \gamma_{m}, \gamma_{t}$ are, respectively, carrier, market, and year fixed-effects. To isolate the effect of common subcontracting on prices, we also control for two variables: the share of passengers transported by regional carriers and multimarket contact. These variables respectively account for factors associated with the costs of using regional carriers, and forbearance among national airlines in regard to multimarket contact.\footnote{Besides the papers mentioned earlier, there is extensive research that shows that multimarket contact leads to higher prices and lower quality, see \cite{Busse2004, ChicuZiebarth2013, MolnarVioliZhou2013, Schmitt2018, LinnMcCarthy2019} and \cite{EizenbergShilianBlanga2023}.}

\subsection{Identification\label{section:identification}}
The main threat to identifying the coefficients in \eqref{eq:model_mmccsc} is that the first three variables are likely endogenous, as we explain next. 
To address that concern, we propose to use instrumental variables. 
Here we introduce three sets of instrumental variables. The first two sets rely on variations in the airlines' network structures, and the third relies on extreme weather patterns across airports.

First, we consider common subcontracting and regional share. 
Omitted variables may affect prices, the level of common subcontracting, and the regional share. In particular, we are concerned about unobserved costs of transporting passengers and mergers among national airlines that may differ across regional and major carriers. 
As mentioned earlier, regional carriers use smaller aircraft, the so-called ``regional jets," and have lower wage bills. 
So the unobserved cost of transporting passengers may be positively correlated with prices and negatively correlated with common subcontracting. Common subcontracting may also (mechanically) change with mergers among national airlines, and in so far as these mergers affect prices, it may result in omitted variable bias.  
For a similar reason, unobserved costs can affect major airlines' subcontracting decisions and, therefore, the regional share.

To address this concern, we use information about the network sizes of \textit{competitors'} regional airlines as our first set of instrumental variables. 
The key idea here is that a major airline's competitors' regional networks in a market affect common subcontracting in that market but do not directly affect the prices charged by that major airline. 
We measure the network size of a regional airline to be the total number of airports that the regional airline operates in a given year. 
To construct the instrumental variable that measures competitors' regional networks, we first take the average network size of all the regionals a major uses in a given market. Then, for each major in the market, we aggregate the average regional network size of the competitors. We refer to this set of IVs as the ``Regional Network IV."

The next set of variables is based on the work by \cite{ForbesLederman2009}, who shows that weather conditions at an airport affect the decisions of major airlines to subcontract operations to regional airlines in a given flight segment.
At the market-major level, we take the most extreme value of four weather condition variables (maximum precipitation, snowfall, snow depth, and lowest minimum temperature) of all airports used along the routes used by the major airline to serve the market. 
We maintain that conditional on our control variables in the pricing regression, including the fixed effects, the weather does not directly affect prices since there should be no ``weather surcharge." 

We use two sets of excluded weather variables: one for common subcontracting and one for regional share. For common subcontracting, we use four measures of the most extreme weather conditions that a major airline's \emph{competitors} encounter in the routes they use to serve a market. Airlines use different routes to serve the same market, which induces variation in the weather across competitors. Common subcontracting is a function of the regional usage by an airline's competitors in a market. So any excluded variables that affect competitors' use of regionals (such as extreme weather) will correlate with common subcontracting. On the other hand, for regional share, we use four measures of the most extreme weather conditions that a given major airline encounters while serving a market as excluded variables. We refer to this set of IVs as ``Weather IVs" and provide additional details in Appendix \ref{section:weather}. 

Finally, the key challenge in identifying the price effect of multimarket contact is that it may be correlated with unobserved market profitability, as discussed in \cite{CilibertoWilliams2014}. Specifically, an airline is more likely to enter a market with a higher expected profit and charge a higher price than in a market with a lower expected profit. Such markets will, therefore, also have higher multimarket contact. 

To address this concern, we rely on the fact that multimarket contact depends on market structure, which in turn depends on the networks of the airlines \citep[e.g.,][]{Berry1992, CilibertoMurryTamer2021}. Therefore, we use instrumental variables that relate to the larger network decisions of major airlines' competitors. 
For each market, we construct a vector of the number of markets that major airlines operating in the market serve both out of the origin and destination airport associated with that market. So, the size of this vector is twice the number of major airlines operating in that market. 
Then, from this vector, for each airline, we consider the vector corresponding to the airline's competitors. This vector captures information about major airlines' competitors' larger networks, which are fixed in the short run and thus affect multimarket contact but not directly the price. 
We label these two sets of excluded variables ``Network IVs." These Network IVs differ from network controls in that the latter are defined for a major carrier, whereas the IVs capture the network size measures of that carrier's market competitors. 

Finally, there may be a concern about the correlation between mergers and common subcontracting. Mergers affect common subcontracting via the market structures. So, mergers may indirectly affect prices through common subcontracting. They may also have \emph{direct} price effects because they change the market structure, hence the ability to exercise market power, and, possibly, because they may reduce costs through efficiency gains. 
However, in our sample, common subcontracting \textit{keeps varying} over time and across markets, whereas mergers are one-time events. So, we rely on that continued variation in common subcontracting and its correlated variation in the IVs for estimation. Besides, the increase in common subcontracting is a more recent phenomenon (Figure \ref{fig:csc_mmc}) than the mergers, suggesting that we can identify the effect of common subcontracting on prices.   

\subsection{Estimation Results}
Column (1) of Table \ref{table:TWFE} presents the fixed effect estimates of the first three coefficients in Equation (\ref{eq:model_mmccsc}).
The coefficient of common subcontracting is imprecisely estimated, and the magnitude is economically insignificant.
We also find that regional share is positively associated with the price, while we expect it to be negative. We find that multimarket contact is associated with higher prices, consistent with previous work on the price effect of multimarket contact. The coefficient of multimarket contact is equal to 0.0135, and it is statistically and economically important. In particular, if we move from a market at the 25th percentile of multimarket contact to one at the 75th percentile, this interquartile range (IQR) move increases multimarket contact by $3.476$, which in turn is associated with a 4.69\% increase in average price. For comparison, in the largest 1,000 markets from 1984 to 1988, \cite{EvansKessides1994} find that an IQR increase in multimarket contact increases prices by $5.1\%$.

\begin{table}[t!]
\caption{\bf{Common Subcontracting, Prices and Quantities}}
\vspace{0.35em}
\hspace{-0.5in}\scalebox{0.8}{\begin{tabular}{lcccccccc} \toprule
{Variables} &  (1) & (2) & (3) &  (4) & (5) & (6)&(7)&(8)\\
 &  $\ln P$ & $\ln P$ & $\ln P$ &  $\ln Q$ & $\ln Q$ & $\ln P$&$\ln P$&$\ln P$\\
\midrule
CSC  & -0.00253  & 0.0834*** & 0.0919***&-0.3442***&-0.8102*** &  0.107***        & 0.108***        & 0.0883***\\
 & (0.00290)  & [0.00534] & [0.0110] & [0.0271] &[0.0039]&[0.00461] & [0.00555]  & [0.00668]\\
{Regional Share}   & 0.0291***  & -0.208*** & -0.126***  &0.1081 &0.2592***& -0.214*** & -0.241*** & -0.187*  \\
 & (0.00175)  & [0.0180] & [0.0195] &[0.103] &[0.1090]& [0.0171] & [.0182] & [0.0197]\\
MMC & 0.0135***  & 0.0231*** & 0.0335*** & 0.0054**&-0.0075*** & 0.0208*** & 0.0230*** & 0.0181***  \\
 &  (0.000397)  & [0.000493] & [0.000566] & [0.0027] &[0.0029]&[0.000509] & [0.000493] & [0.000592]\\
 Regional HHI &&&&&&&& 0.131*** \\
&&&&&&&& [0.00694]\\
   CSC 2004-2012   &  &  & 0.0112&&0.5304***&& \\
   &  &  & [0.0103] &&[0.0308]&&\\
  CSC Post-2012    &  &  & 0.0278***&&0.4102***&&\\
   &  &  & [0.0101]&&[0.0308]&&  \\
Regional Share 2004-2012  &  &  & -0.0278*** &&-0.2208***&&\\
  &  &  & [0.00586]&&[0.0211]&&\\
 Regional Share Post-2012   &  &  & -0.0490***&&-0.2961***&&\\
  &  &  & [0.00583] &&[0.0219]&& \\
  MMC Post-2010  &  &  & -0.0150***&&0.0205***&& \\
    &  &  & [0.000371] &&[0.0017]&&\\

\midrule
{\bf First-Stage Results} && & && &&&\\
 &&  & &&&\\

\emph{Residuals:} CSC  &  & -0.106*** & -0.118*** &0.332*** &0.3668***& -0.164*** & -0.133*** & -0.101***\\
 &  & [0.00596] & [0.00598]&[0.0299] &[0.0302]& [0.00510]         &   [0.00602]              & [0.00727] \\
{Regional Share}  &  & 0.240*** & 0.188***&-0.334*** &-0.2622***& 0.251*** & 0.274*** & 0.212***\\
&  & [0.0181] & [0.0186]& [0.1043] &[0.1056]&[0.0172] & [0.0182] & [0.00198]\\
MMC &   & -0.0155*** & -0.0144***  & -0.066*** &-0.0691***&-0.0111*** & -0.0153*** & -0.0111***\\
 & & [0.000493] & [0.000595] &[0.0032] &[0.0032]& [0.000611] & [0.000591] & [0.000691] \\
\hline\\ 
\emph{F-stats (IVs):}  CSC&   & 1,240.34 & 1,240.34  & 1,240.34&1,240.34 &2,369.70 & 754.94 & 1,240.34 \\
 Regional Share&   & 649.96 & 649.96  & 649.96 &649.96&649.96 & 649.96 & 649.96\\
MMC&   & 5,350.11 & 5,350.11  & 5,350.11 &5,350.11&5,350.11 & 5,350.11 & 5,350.11 \\
Regional HHI&   & &   &&  &&  & 1,079.66\\
\midrule
Control Functions&\xmark & \checkmark  & \checkmark &\checkmark&\checkmark&\checkmark&\checkmark&\checkmark \\
CSC (benchmark) &\checkmark & \checkmark &\checkmark&\checkmark &\checkmark&\xmark & \xmark &\checkmark\\
CSC (count) &\xmark & \xmark& \xmark& \xmark&\xmark &\checkmark&\xmark&\xmark\\
CSC (weighted)&\xmark & \xmark& \xmark&\xmark &\xmark&\xmark&\checkmark&\xmark\\
\midrule
Observations & 1,000,180 & 1,000,180 & 1,000,180&1,000,180 &1,000,180& 1,000,180 & 1,000,180 & 802,265 \\
Number of markets & 21,390 & 21,390 & 21,390 & 21,390&21,390 &21,390 & 21,390 & 20,815\\
R$^2$ & 0.092 & 0.094 & 0.096  & 0.29&0.29 &0.094 & 0.093 & 0.102 \\
\bottomrule
\end{tabular}}\label{table:TWFE}
\begin{figurenotes}
The table presents the estimates of price regression \eqref{eq:model_mmccsc} in columns (1)-(3) and (6)-(8).  Columns (4)-(5) present the estimates of quantity regression where the dependent variable in \eqref{eq:model_mmccsc} is replaced with the log of passenger traffic (number of passengers transported). Here, CSC stands for common subcontracting  MMC stands for multimarket contact. Estimates in columns (2)-(8) use control functions. When we estimate the first-stage residual for multimarket contact, we omit the regional-specific Weather IVs and Regional Network IVs. Market fixed effects, network IVs, weather IVs, and reg network IVs were used in all regressions. Other control variables included in all specifications but whose coefficients are not reported include dummy variables for year-quarter, markets, and carriers, and controls for network size measured by origin and destination networks. Columns (6)--(8) use the same specification as column (2), except column (6) uses common subcontracting (count), column (7) uses common subcontracting (weighted), and column (8) uses the baseline definition of common subcontracting but includes market-specific Herfindahl-Hirschman Index of regional carriers (Regional HHI) based on the share of passengers transported. For column (2), robust standard errors are in parenthesis; for the rest, bootstrapped standard errors based on 1,000 samples are in square brackets.  \end{figurenotes}
\end{table}

The results in column (1) are likely biased due to the endogeneity of all three variables. 
To address this concern, column (2) of Table \ref{table:TWFE} presents estimates using a control function approach using three  instrumental variables: ``Regional Network IVs," ``Weather IVs," and ``Network IVs."
We also present coefficients for the first-stage residuals of all three variables. 

We find that the coefficient for common subcontracting increases to 0.0834, and the coefficient is precisely estimated. This result is consistent with the negative coefficients for the first-stage residual for common subcontracting. Thus, common subcontracting is associated with higher prices. In particular, an IQR increase in common subcontracting from 0 to 0.214 increases average prices by 1.8\%.\footnote{The negative bias of the estimate without the instrumental variables is consistent with the findings of \cite{CarltonIsraelMacSwainOrlov2019}, who find that mergers are pro-competitive.}

We also find that the regional share \textit{lowers} the price, which confirms the reasoning that relying on regional airlines in a market should lower costs. The coefficient of regional share is precisely estimated at $-0.208$, so an IQR increase in regional share lowers prices by $3.9\%$. 

The estimate for multimarket contact at 0.0231
implies that an IQR increase in multimarket contact by $3.476$ increases the average price by 8\%. For comparison, \cite{CilibertoWilliams2014}, find that an IQR increase in multimarket contact increases prices by 6.5\%.

We also explore whether, given the structural changes we observe in the industry (Figures  \ref{fig:regionalusage}, \ref{fig:regionalusagetrend}, \ref{fig:reg_share_average} and  \ref{fig:csc_mmc}), the price effects have changed over time. 

To capture the time trend, we define categorical variables associated with the periods where the variables showed different patterns. For common subcontracting and regional share, we consider the period before (including) 2003, between 2004 and 2012, and after 2012. In the case of multimarket contact, we consider the period before and after 2010. We then interact these categorial variables with the three variables of interest and include them as additional regressors to \eqref{eq:model_mmccsc}.

The estimates using all the instrumental variables are presented in column (3) of Table \ref{table:TWFE}. The coefficient estimate of Common Subcontracting 2004-2012 is the estimated difference between the baseline pre-2004 estimate of Common Subcontracting and the effect from 2004-2012. Similarly, the coefficient estimate of Multimarket Contact Post-2010 can be interpreted as the change in the effect of multimarket contact relative to the coefficient estimate of Multimarket Contact. The coefficient estimates are precisely estimated, except for the difference between the common subcontracting estimate pre-2004 and the estimate in the period 2004-2012, which is consistent with the generally low level of regional share in those periods. Thus, there is no difference in the effect of common subcontracting in 2004-2012 relative to pre-2004. 

We find that the price effect of common subcontracting is larger in the post-2012 period than before, and the negative price effect of regional share is stronger after 2004, suggesting that the cost savings have increased. However, we find that the price effect of a multimarket increase is smaller in later years when the absolute level of multimarket contact is larger. 

For ease of comparison, Table \ref{table:trend_effect} presents the price effect of increasing common subcontracting, regional-share, and multimarket contact by their specific IQR, separately for periods implied by the estimates in column (3) of Table \ref{table:TWFE}.

\begin{table}[t!]
\centering
\caption{\bf{Price Effects over Time}\label{table:trend_effect}}
\scalebox{1}{ \begin{tabular}{cccccc}
       \toprule
          &    Years   &  &Common Subcontracting& Regional-Share &Multimarket Contact \\
          \midrule
    & Pre-2004  &       & 0.06\% & 0\%& \\
          &  2004-2012  &       & 1.72\% & -3.92\%&  \\
          &Post-2012  &       & 5.07\% & -7.42\%&  \\
           & Pre-2010 &     &    && 7.35\% \\
           & Post-2010 &  &  && 9.13\% \\
          \bottomrule
\end{tabular}%

}
\begin{figurenotes}
Price effects of an IQR increase in common subcontracting, regional share, and multimarket contact, as implied by the estimates in column (3) Table \ref{table:TWFE}. Pre-2004 means 1996 to 2003, post-2004 means 2004 to 2012, and post-2012 means 2013 to 2016. Similarly, pre-2010 means from 1996 to 2009, and post-2010 means 2010 to 2016. 
\end{figurenotes}
\label{table:marginaleffectsMMC}
\end{table}%

These estimates suggest that if we increased the common subcontracting by its IQR, it would not have affected prices pre-2004 because common subcontracting was negligible, with IQR effectively zero. In 2004-2012, if we increase common subcontracting by its IQR, the price would increase by  $1.72\%$, but by post-2012, the price would increase by 5.07\%. 
Thus, the potential anticompetitive effect of common subcontracting has recently become stronger. 

The decrease in prices due to regional share shows a similar trend as common subcontracting. 
In particular, the estimates suggest that an IQR increase in regional share would have had no effect on price pre-2004, but in the years 2004-2012 and post-2012, the price would have decreased by $3.92\%$ and  $7.42\%$, respectively. 

The growth rate of the price effect for multimarket contact is lower than common subcontracting. 
Our estimates suggest that an IQR increase in multimarket contact leads to a 7.35\% pre-2010 and 9.13\% post-2010. 
Thus the increase in price effects for multimarket contacts is relatively smaller than that for common subcontracting, which is consistent with the difference is the increase in the level of common subcontracting and multimarket contact (see Figure \ref{fig:csc_mmc}). As mentioned before, the growth in common subcontracting is due to the use of regional airlines is a relatively recent phenomenon, fueled by introduction of new regional jets, rising consolidation among regional airlines and mergers among major airlines.
   
 Finally, we note that a difficulty in interpreting the estimated increase in price as an anticompetitive effect of common subcontracting is that prices could be higher because of higher quality air transport services. If so, we expect common subcontracting to either increase or not affect passenger traffic (number of passengers transported). 
However, we find that common subcontracting decreases traffic, measured by log of passengers transported. 
In particular, we estimate the specifications in columns (2) and (3) of Table \ref{table:TWFE} with traffic as the dependent variable, and the estimates are in columns (4) and (5) of Table \ref{table:TWFE}. We find that an IQR increase in common subcontracting leads to a 7.4\% decrease in traffic. Furthermore, before 2004, an IQR increase in common subcontracting decreases traffic by 0.6\% before 2004, by 4.7\% between 2004 and 2012, by 16.9\% post 2012.

\section{Robustness\label{section:robustness}}
 In this section, we investigate whether the results in Section \ref{section:result} are sensitive to how we define common subcontracting and the concentration in the regional markets.

\subsection{Alternative Measures of Common Subcontracting}

We consider two alternative measures to assess whether our results are sensitive to our particular method of constructing common subcontracting. 
In the first one, we use only the number of overlapping regional airlines instead of weighting by the fraction of passengers flown by regional airlines. In the second one, we use the market shares of the major airlines to construct a weighted measure of overlap in the use of regional airlines. 

The first alternative measure, which we refer to as ``\textit{common subcontracting (count)}," does not depend on the share of passengers transported by a regional airline but on the number of overlaps across major airlines created by regional usage. This measure is defined as follows. We determine how many unidirectional major airline pairs are present in each market and each year. This number is two for markets with only two major airlines; for markets with three, this number is six, and likewise for other markets. Next, we determine whether each unidirectional pair has overlapping regional usage and sum the number of pairs with overlapping regional usage. Then we define the common subcontracting (count) measure as the ratio of the number of unidirectional major pairs with overlapping regional usage to the number of total unidirectional major pairs. 

For example, suppose AA, DL, and UA are in a market. AA uses a regional airline in this market, and DL uses that airline in another market. No other regionals are used. Then there is one unidirectional major pair for which there is overlap out of six total unidirectional pairs in the market. So the common subcontracting (count) is $1/6=.167$. 
Like the baseline measure, this count-based measure also takes on values between $0$ and $1$. 
In our sample, the mean of common subcontracting (count) is $0.238$ with a high coefficient of variation of $1.26$.

We re-estimate the model specification for column (2) in Table \ref{table:TWFE} using the control function method with common subcontracting (count) in place of our preferred common subcontracting (baseline) measure. All other variables remain the same. The results are presented in column (6) of Table \ref{table:TWFE}. 
The estimates are qualitatively the same as in column (2).

Next, we explore the effects of weighting the overlap across regional airlines by the market share of a major airline. 
We refer to this measure as ``\textit{common subcontracting (weighted)}." 
For instance, consider the example presented in Table \ref{table:exampleofcscpart1}. 
AA uses a regional airline, OO, to transport a $0.667$ fraction of its passengers. DL, the other major in the example market, used OO in another market in the same period. 
To construct common subcontracting (weighted), we multiply the fraction of passengers for which AA subcontracts operations to an overlapping regional airline and AA's market share. In this example market, AA's market share is $.429$, and DL's is $.571$, so the weighted common subcontracting is $(.667\times.429 + 0 \times .571)/2 = .143$.
The overall mean value of common subcontracting (weighted) at $0.163$ is slightly smaller than the mean of common subcontracting (count) but has a higher coefficient of variation of $1.62$.

We re-estimate the model using the control function method with weighted common subcontracting. 
The estimates in column (7) of Table \ref{table:TWFE} are similar to those in column (2).

\subsection{Concentration among Regional Airlines}

Our estimates suggest that the overlapping usage of regional airlines raises the price, especially in the later years in the sample.
An alternative explanation for our results could be that concentration among regional airlines is driving this association: regional airlines with more bargaining power can extract higher fees from legacy airlines and drive up costs and, in turn, prices. For example, we observe that the largest regional airline in our sample, ExpressJet, served at least one segment of 55\% of total markets in 2007, and it served 70.5\% in 2016. This increase could either be due to the increasing usage of regionals (see Figure \ref{fig:regionalusagetrend}) or be the result of M\&A activity and consolidation as suggested by Table \ref{fig:chronology}.

To address this question, we calculated what we refer to as \emph{Regional HHI}. To estimate the underlying ``market'' shares for this measure, we divided the number of passengers transported by a given regional airline by the total number of passengers in a market served by all regionals. Then we summed the square of those share terms across all regional airlines active in each market. We find that the interquartile range of concentration stays relatively constant over the sample and that the median concentration decreases. 

To assess the explanatory power of regional concentration, we re-estimate the model with ``{Regional HHI}" as an additional covariate. 
The HHI and prices are determined simultaneously in equilibrium, so we treat regional HHI as an endogenous variable and use the Regional Network IVs and Weather IVs as excluded variables.  
The estimation results are in column (8) of Table \ref{table:TWFE}.
While prices are positively correlated with market concentration among regional carriers, comparing the estimates in column (8) with (2), we find that the estimated coefficients on common subcontracting are similar to our benchmark results.

\section{Conclusion}

In this paper, we investigate a new channel through which firms can charge higher prices: sharing subcontractors across markets, i.e., common subcontracting. 
We find that common subcontracting leads to higher prices and lower air traffic. 
We also find substantial benefits to using regional airlines because they lead to lower prices. 
However, the increasing extent of common subcontracting in recent years has significantly reduced those benefits. 

Thus, there is a need for a nuanced approach to understanding the vertical structure in the airline industry and how it affects competition.
One hypothesis is that sharing regional airlines across markets allows major airlines to improve information about each other's costs, which helps firms collude. However, testing this hypothesis is beyond the scope of our paper. 
As such, there are still unanswered questions about what mechanism drives these results and how they generalize to other industries and subcontracting structures. Based on a rich literature that suggests that \emph{information} and communication help collusion \citep{Roberts1985, Shapiro1986,  Kandori1992, KuhnVives1994, AtheyBagwell2001, Rahman2014, AwayaKrishna2016, PiccoloThal2012, AryalCilibertoLeyden2022}, we conjecture that common subcontracting softens competition by
providing major carriers with more information about each others' costs.

Studying collusive behavior among firms is central to economics \cite[e.g.,][]{Harrington2017, Porter2005, Kaplow2013, AskerNocke2021}.  
Nevertheless, determining how firms reach a collusive setting is challenging.
Thus, the literature has approached collusion on a case-by-case basis, and we are cognizant that we raise more questions about the role of common subcontracting in collusion than provide answers. We hope our results will spark more interest in understanding of the role of subcontractors on market outcomes.

\bibliographystyle{econometrica} 
\bibliography{bibliography.bib}
\setcounter{section}{0}
\setcounter{equation}{0}
\setcounter{figure}{0}
\setcounter{table}{0}
\renewcommand{\thesection}{A.\arabic{section}}
\renewcommand{\theequation}{A.\arabic{equation}}
\renewcommand\thefigure{\thesection.\arabic{figure}}
\renewcommand\thetable{\thesection.\arabic{table}}

\small
\section{Appendix: Sample Construction\label{section:sample}} 

The main data are from the domestic Origin and Destination Survey (DB1B), which is a 10\% sample of airline tickets from all reporting carriers. This dataset has three versions.
First, we start with the DB1B Coupon version, which is the basis of our analysis. 
This dataset provides coupon-specific information for each domestic itinerary of the Origin and Destination Survey, such as the name of the operating carrier, origin and destination airports, and number of passengers. 
We use this dataset to construct the chain of routes each ticketing carrier uses to serve a market, which, as defined in the paper, is a unidirectional trip between two airports, irrespective of intermediate transfer points. This dataset also identifies the regional carriers for each nonstop segment (route). From this dataset, we keep only domestic flights, tickets sold by domestic carriers, flights between the 48 contiguous United States, and flights that require at most six coupons.

Second, we merge the Coupon version with the DB1B Market version, which contains information on the directional market characteristics for each Origin-Destination domestic itinerary. We exclude bulk fare tickets and tickets with more than three coupons in either direction.
Third, we merge in the DB1B Ticket version containing information on the reporting carrier, itinerary fare, number of passengers, originating airport, roundtrip indicator, and miles flown. We exclude tickets whose fare credibility is questioned by the DOT.

We further exclude tickets that are neither one-way nor roundtrip travel or include travel on more than one airline on a directional trip (known as interline tickets), here identified by whether there was a change in the ticket carrier for the ticket.

Next, we follow the approach in \cite{Borenstein1989} and \cite{CilibertoWilliams2014} and consider a roundtrip ticket as two directional trips on the market, and the fare for each directional trip is half of the roundtrip fare. A one-way ticket is a one-directional trip. From the final sample, we exclude tickets with fares less than \$20 in the top and bottom one percentile of the year-quarter fare distribution and for which the fare per mile (the yield)  in the top and bottom one percentile of the year-quarter yield distribution.

\section{Appendix: Weather\label{section:weather}} 
In this section, we discuss the construction of our Weather IVs using airport-level weather conditions data from the Daily Global Historical Climatology Network \citep{Menne2012a, Menne2012b}. We use four of the ``core" variables in the data: precipitation (in tenths of mm), snowfall (mm), snow depth (mm), and minimum temperature (tenths of degrees celsius). These variables are recorded for each airport at the daily level. For our purposes, we average across all days in a quarter to arrive at average quarterly levels. Summary statistics across airports and years for these variables are presented in the Table (\ref{tab:weather_sum}).
\begin{table}[ht!]
    \centering
    \caption{\bf Summary Statistics of Weather Variables}
    \label{tab:weather_sum}
    \scalebox{0.9}{\begin{tabular}{c|c|c|c}
    \toprule
        Variable & Mean & Median & Std. Dev. \\ \hline
        Precipitation (tenths of mm) & 26.46 & 24.84 & 16.70 \\
        Snowfall (mm) & 1.30 &  0.00 & 3.52 \\
        Snow Depth (mm) & 5.77 & 0.00 & 23.96 \\
        Minimum Temp (tenths of degrees celsius) & 95.94 & 101.04 & 83.86 \\ \bottomrule
    \end{tabular}}
    \begin{figurenotes}
    Summary table of the weather variables.
    \end{figurenotes}
\end{table}%
To construct the weather instruments we implement the following steps. First, we consider the quarterly average weather variables for each airport that is used by a major to serve a market in the given year. For example, if AA serves the market CHO-DFW by using the route CHO-CLT-DFW, then we consider the weather variables at all three airports: CHO, CLT, and DFW. At the market-major-year quarter level, for each variable listed above, we take the most extreme value (maximum in the case of precipitation, snowfall, and snow depth; minimum for minimum temperature) of all airports used along the routes the major uses to serve the market. 
For instance, if DFW has the worst precipitation of CHO, CLT, and DFW, we take DFW's average rain value. This gives us a set of four variables for each major in a market-year that record the average weather conditions of the ``worst" airport used to serve the market for each of the weather variables. 
Then for each major in a market, we sum its competitors' extreme weather values in that market to determine the weather IVs used for common subcontracting, as common subcontracting is a function of competitors' regional usage in a market. For regional share of airline $j$, we use the four most extreme values of weather encountered by $j$ itself in the market, as weather conditions are a factor that affect airlines' subcontracting choices.
 \end{document}